\documentclass[floatfix,aps,twocolumn,prb,superscriptaddress]{revtex4-1}

\usepackage{amssymb}
\usepackage{amsmath}
\usepackage[dvips]{graphicx}
\usepackage{bm}
\usepackage{float}
\usepackage{color}

\newcommand{\Ef}{E_{\text{F}}}
\newcommand{\kt}{k_{\text{B}}T}
\newcommand{\be}{\begin{equation}}
\newcommand{\ee}{\end{equation}}
\newcommand{\bea}{\begin{eqnarray}}
\newcommand{\eea}{\end{eqnarray}}

\newcommand{\ND}{\mathcal{N}_D}

\newcommand{\eps}{\epsilon}
\newcommand{\td}{t_{dD}}
\newcommand{\meV}{\,\mathrm{meV}}

\begin{document}

\title{Transport through side-coupled double quantum dots: from weak to strong interdot coupling}

\author{D. Y. Baines}
\affiliation{Institut N\'eel, CNRS and Universit\'e Joseph Fourier, 38042 Grenoble, France}
\author{T. Meunier}
\affiliation{Institut N\'eel, CNRS and Universit\'e Joseph Fourier, 38042 Grenoble, France}
\author{D. Mailly}
\affiliation{Laboratoire de Photonique et Nanostructures, CNRS, route de Nozay, 91460 Marcoussis, France}
\author{A. D. Wieck}
\affiliation{Lehrstuhl f\"ur Angewandte Festk\"orperphysik, Ruhr-Universit\"at, Universit\"atsstra\ss e 150, 44780 Bochum, Germany}
\author{C. B\"auerle}
\affiliation{Institut N\'eel, CNRS and Universit\'e Joseph Fourier, 38042 Grenoble, France}
\author{L. Saminadayar}
\affiliation{Institut N\'eel, CNRS and Universit\'e Joseph Fourier, 38042 Grenoble, France}
\author{Pablo S. Cornaglia}
\affiliation{Centro At{\'o}mico Bariloche and Instituto Balseiro,
CNEA, 8400 Bariloche, Argentina}
\affiliation{Consejo Nacional de Investigaciones Cient\'{\i}ficas y T\'ecnicas (CONICET), Argentina}
\author{Gonzalo Usaj}
\affiliation{Centro At{\'o}mico Bariloche and Instituto Balseiro,
CNEA, 8400 Bariloche, Argentina}
\affiliation{Consejo Nacional de Investigaciones Cient\'{\i}ficas y T\'ecnicas (CONICET), Argentina}
\author{C. A. Balseiro}
\affiliation{Centro At{\'o}mico Bariloche and Instituto Balseiro,
CNEA, 8400 Bariloche, Argentina}
\affiliation{Consejo Nacional de Investigaciones Cient\'{\i}ficas y T\'ecnicas (CONICET), Argentina}
\author{D. Feinberg}
\affiliation{Institut N\'eel, CNRS and Universit\'e Joseph Fourier, 38042 Grenoble, France}
\date{\today}

\begin{abstract}
We report low-temperature transport measurements through a double quantum dot device in a configuration where one of the quantum dots is coupled directly to the source and drain electrodes, and a second (side-coupled) quantum dot interacts electrostatically and via tunneling to the first one. 
As the interdot coupling increases, a crossover from weak to strong interdot tunneling is observed in the charge stability diagrams that present a complex pattern with mergings and apparent crossings of Coulomb blockade peaks. While the weak coupling regime can be understood by considering a single level on each dot, in the intermediate and strong coupling regimes, the multi-level nature of the quantum dots needs to be taken into account. 
Surprisingly, both in the strong and weak coupling regimes, the double quantum dot states are mainly localized on each dot for most values of the parameters. 
Only in an intermediate coupling regime the device presents a single dot-like molecular behavior as the molecular wavefunctions weight is evenly distributed between the quantum dots.
At temperatures larger than the interdot coupling energy scale, a loss of coherence of the molecular states is observed.
\end{abstract}

%\pacs{03.65.w, 03.67.Mn, 42.50.Dv}

\maketitle

\section{Introduction}
Double quantum dot (DQD) devices have been the object of numerous experimental\cite{Hofmann1995,Wiel2002,Chen2004,Potok2007} and theoretical \cite{Loss1998,DasSarma2000,Boese2002} studies due to their potential applications in both classical and quantum computing,\cite{Loss1998,DasSarma2000,Koppens2006,Hayashi2003,Hanson2007} and also because of their usefulness as model systems to study the physics of strongly correlated electrons.\cite{Georges1999,Vojta2002,Potok2007,PhysRevB.71.075305,PhysRevLett.97.166802,0953-8984-23-24-243202} 

The transport signatures of these devices depend strongly on their topology and on the geometry of the quantum dots, which determines their energy level spacings and charging energies.\cite{GrabertD92,Kastner92,Beenakker1991,MesoTran97,*KouwenetalRev97,AleinerBG02,Alhassid00} 
In the so-called side-coupled configuration, were only one of the quantum dots is coupled to the electrodes, a rich variety of correlated phenomena has been predicted,\cite{Boese2002,PhysRevB.71.075305,Zitko2006,*Zitko2010,Karrasch2006,Chung2008,Cornaglia2011} yet few experiments are available.\cite{Potok2007,Sasaki2009} Signatures of two-channel Kondo physics have been measured for a device with a {\it large} side-coupled quantum dot, \cite{Potok2007} while a two-stage Kondo effect has been proposed for a small (single-level) side-coupled dot in the Kondo regime. \cite{PhysRevB.71.075305}
Furthermore, a device with an intermediate size of the side-coupled dot has been predicted to be a realization of the Kondo box problem. \cite{PhysRevB.73.205325,PhysRevLett.82.2143,Hu2001,Simon2002,Cornaglia2003b,*Cornaglia2002a,Kaul2005,Yoo2005,Bomze2010} 
As we will show in what follows, in the weak quantum dot-electrodes tunneling regime, where the Kondo effect is exponentially suppressed, this type of device allows for a controlled study of the interplay between the interdot tunnel coupling, the temperature and the multiple levels of the quantum dots.  

In this work we measure the electronic transport through a DQD in the side-coupled configuration and characterize the effect of the interdot tunnel coupling and the temperature.
For sufficiently weak interdot tunneling coupling, the charge stability diagrams can be understood within the usual two-level representation.\cite{Wiel2002} As the tunneling coupling increases, however, the device enters a molecular regime where the multi-level nature of the quantum dots needs to be taken into account in order to capture the physics of the low temperature regime.

Numerical simulations based on a simplified multi-level double dot Hamiltonian enable to calculate the conductance through the system in the sequential tunneling regime. We show that a qualitative understanding of the transport properties at low and high temperature can be reached by comparing experimental and numerical data in the energy range where the experiment is carried out.

The rest of the paper is organized as follows: 
In Sec. \ref{sec:exp} we describe the experimental setup. 
In Sec. \ref{sec:tm} we present transport measurements illustrating the weak to strong interdot tunneling crossover and the effect of temperature. 
In Sec. \ref{sec:theory} we present the model and methods for the calculation of the conductance. 
In Sec. \ref{sec:numdis} we present numerical results that reproduce the main features observed experimentally in the crossover from weak to strong interdot tunneling. 
In Sec. \ref{sec:regimes} we characterize numerically and analytically the different interdot tunneling and temperature regimes.
Finally, in Sec. \ref{sec:concl} we present our concluding remarks.

\section{Experiment} \label{sec:exp}
\begin{figure}[t]
\includegraphics[width=3.4in]{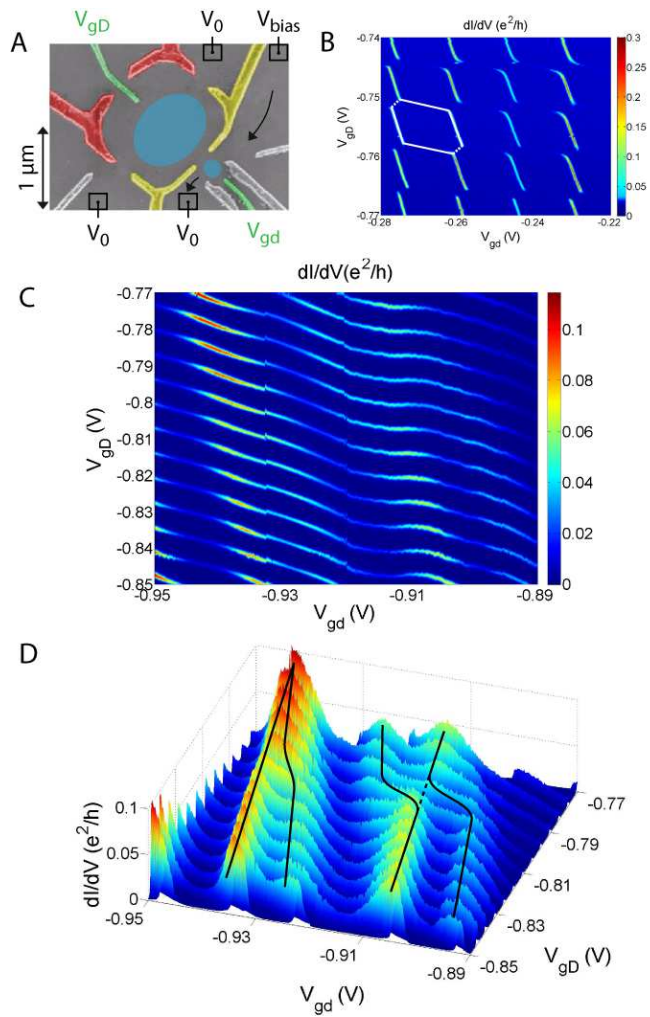}
\caption{(Color online) \textbf{A}: SEM image of the device. The red gates are pushed far in the pinch-off regime. The interdot tunnel coupling is controlled via the yellow gates. All ohmic contacts are set to the same potential (V$_{0}$) except the top small dot lead where the bias voltage $V_\text{bias}$ is applied. The energies and occupancies of each dot are changed with the use of the green plunger gates. \textbf{B}: Weak coupling stability diagram (30 mK). The differential conductance (color scale) is plotted versus the small dot and large dot plunger gate voltages, V$_{gd}$ and V$_{gD}$ respectively. \textbf{C} and \textbf{D}: Low temperature (30 mK) stability diagrams (2D and 3D) with stronger coupling between the dots. At such low temperatures, a complex conductance modulation pattern is found.}
\label{Fig1}
\end{figure}

Our device consists of a double quantum dot defined in a two-dimensional electron gas formed in a GaAs/Al$_x$Ga$_{1-x}$As heterostructure (density 2.4$\times$10$^{11}$ cm$^{-2}$, mobility 1$\times$10$^6$ cm$^2$V$^{-1}$s$^{-1}$). The quantum dots are designed following a side-coupled (or T-shape) configuration where a small quantum dot (500 nm) is connected to electron reservoirs and is side-coupled to a large quantum dot (1500 nm) (Fig.~\ref{Fig1}, A). The bare charging energy and mean level spacing of each dot extracted from non linear measurements in the weak coupling regime are: U$_d$=700 $\mu$eV, $\Delta_d$=150 $\mu$eV and U$_D$=250 $\mu$eV, $\Delta_D$=20 $\mu$eV, for small dot (d) and large dot (D) respectively. The differential conductance $dI/dV$ of the double dot system is measured by applying a DC and AC (11 Hz, 2 $\mu$V) voltage excitation on the top small quantum dot lead, as indicated by $V_\text{bias}$ in Fig. \ref{Fig1} (A). In this particular electrical set-up, transport occurs only through the small quantum dot. The side-coupled large quantum dot influences the transport mechanisms via the interdot tunnel coupling that is controlled with the voltages applied on the middle gates. All measurements are performed in a dilution refrigerator with a base electron temperature of 30 mK. The electronic temperature for different fridge temperatures has been calibrated from weak localization measurement realized in earlier experiments.\cite{Yasu,*YasuB}

\section{Transport measurements}\label{sec:tm}
The regime of interest is a strong interdot tunnel coupling regime compared to the weak coupling limit widely studied in lateral quantum dots.\cite{Pothier} In order to emphasize the influence of the increase of the interdot tunnel coupling in our system we make a comparison by means of Fig.~\ref{Fig1} (B, C and D) that shows two experiments performed on the same double dot system at low temperature (30 mK) and with different interdot hopping strengths. Figure~\ref{Fig1} (B) shows the weak coupling limit. The conductance pattern follows a honeycomb lattice. Note that due to the side-coupled configuration, conductance occurs mainly at the degeneracy points of the small dot with the Fermi energy.\cite{Beenakker1991} Nevertheless, detection of current on the degeneracy line of the large dot with the reservoir indicates finite though weak interdot hopping. The position of the charge states of each dot can therefore be identified through the modulation of the conductance on the different degeneracy lines. In other words, a large conductance peak indicates a charge state holding an important weight of the small dot wavefunction and vice versa. Moreover, the small dot-leads hybridization ($\Gamma$) results in very thin degeneracy lines and in a very low conductance in the Coulomb blockade valleys (10$^{-4}$ e$^2$/h). Such a regime of weak tunneling (to the leads and between the dots) can be accounted for in a standard two-level representation.\cite{Wiel2002} The stronger coupling situation is met in Fig.~\ref{Fig1} (C and D). One can notice that the degeneracy lines seen in the stability diagram still appear as thin conductance lines indicating a rather weak coupling to the leads. This point can be confirmed by monitoring the conductance in the Coulomb blockade valleys which is of the order of 10$^{-3}$ e$^2$/h which is still far from the strong coupling limit ($\sim$0.1 e$^2$/h).\cite{vdWiel2000} We will consider therefore the tunnel coupling to the leads as a weak perturbation and concentrate mainly on the effect of the interdot hopping. Concerning the conductance pattern, we observe that it deviates from a honeycomb lattice. From Fig.~\ref{Fig1} (C) one can argue that the effect of enhanced tunneling between the dots is an effective smoothing of the honeycomb structure which one can expect from earlier literature.\cite{Wiel2002} However the nonperiodic modulation of the conductance in each direction of the voltage gate space depicted in Fig.~\ref{Fig1} (D) shows unusual transport features such as the apparent merging and the crossing of peak structures in the $V_{gD}$ direction (black lines). Such conductance features cannot be accounted for in the common two-level representation. To be able to capture the low temperature physics, the multi-level structure of the large dot on the energy range of the interdot tunnel coupling, $t_{dD}>\Delta_D$, has to be considered.

\begin{figure}[!h]
\includegraphics[width=3.4in]{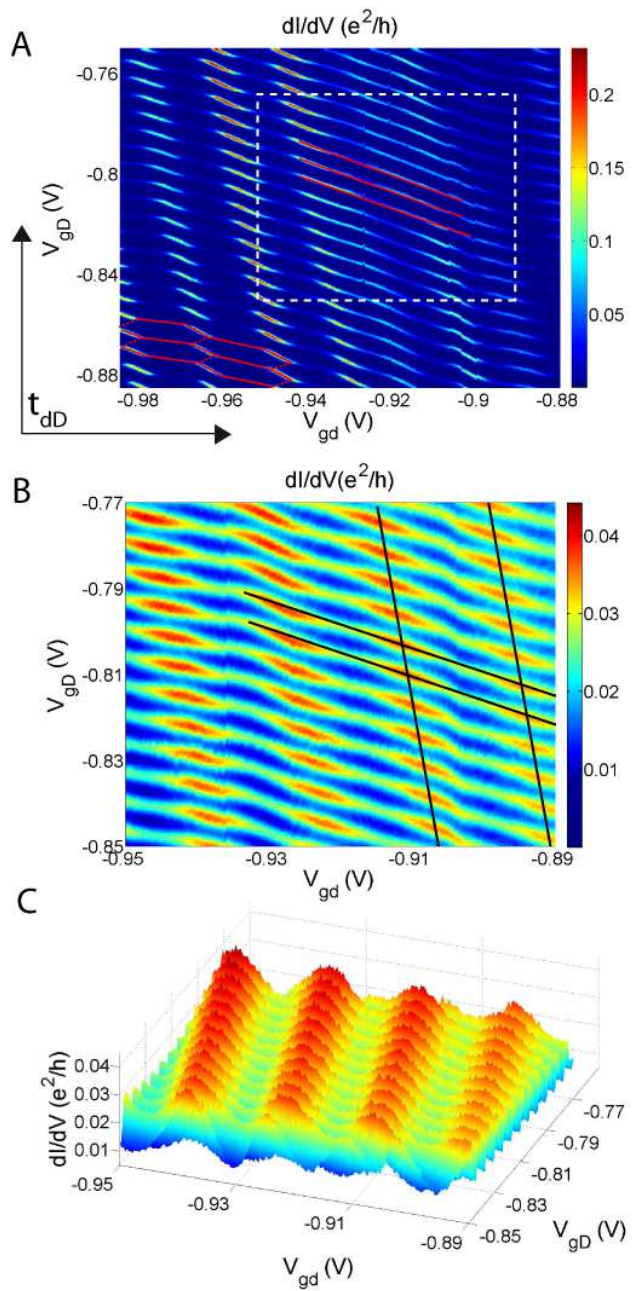}
\caption{(Color online) \textbf{A}: Low temperature stability diagram (30 mK) corresponding to a larger scan than previously shown. The white dashed rectangle corresponds to the gate scan seen in Fig.~\ref{Fig1} B,C. Due to cross-talk between the plunger gates and the middle gates defining the interdot tunneling two different regimes can be identified in the diagram. As tunneling increases we go from a two-level system behavior to a multi-level system behavior. \textbf{B}: Color plot of the high temperature stability diagram monitored at 500 mK in the same plunger gate voltages range made initially (Fig~\ref{Fig1} B,C). \textbf{C}: Three-dimensional representation of the above diagram. At high temperature, the conductance follows a periodic pattern. One can clearly identify the addition of electrons in one quantum dot or the other.}
\label{Fig2}
\end{figure}

In order to justify the use of a multi-level representation of the hybridization between both quantum dots we make use of Fig.~\ref{Fig2} (A). The depicted low-temperature stability diagram represents a larger gate voltage scan centered around the same region of parameters used in the previous low temperature measurements (white dashed rectangle). From Fig.~\ref{Fig2} (A), it appears that the area scanned in Fig.~\ref{Fig1} (C, D) corresponds to a crossover region between two different regimes. On the bottom left side of Fig.~\ref{Fig2} (A) one can identify smoothed honeycomb cells (red cells). By studying the deviation of these cells from pure honeycomb cells, we can extract a rough estimation of the interdot tunnel coupling, that is to say $t_{dD}\sim 30\mu eV$. In this limit, $t_{dD}\sim \Delta_D$, the experimental data point towards a two-level system behavior, hence the apparent honeycomb cells. However by depolarising both plunger gates one can note through the straightening of the degeneracy lines (red lines) that the hopping term increases. This feature can be easily understood due to crosstalk between the different gates. Once the regime $t_{dD}>\Delta_D$ is reached, hybridization involves multiple energy levels in dot $D$. As a result the molecular addition spectrum~\cite{Bruus} gains in complexity which we expect to be reflected in the conductance through the device. A proper calculation will be presented below to illustrate this point.

Before going into the core of the multi-level double dot model, it is interesting to address the question of the evolution of the transport properties as temperature is increased.  Experimental data shown in Fig.~\ref{Fig2} (B and C) indicate that at high temperature, 500 mK, the strong irregularities found at low temperature are completely washed out and a periodic pattern is recovered (Fig.~\ref{Fig2} B, black lines). Whereas at low temperature one cannot interpret the conductance via the filling of both dots independently, in the high temperature limit the standard picture can be applied. Periodic oscillations of the conductance as a function of both plunger gate voltages enable to keep track of the addition of electrons in each quantum dot separately. Therefore by heating the device, the coherence of the molecular eigenstates is broken and a suitable description is found by simply thinking in terms of the occupancy states in dots d and D. More precisely, by bringing thermal energy to the system, the Fermi distribution of the metallic leads is broadened which leads to a larger effective conduction window. We argue that once $\kt>t_{dD}$, the conductance measured through the double quantum dot represents an average over several molecular levels lying in the conduction window which results in the loss of coherence of molecular states and leads to a regular stability diagram. The picture at low temperature becomes clearer now. As we will show below, for $t_{dD}>\Delta_D>\kt$ and at the degeneracy with the leads, the addition of an electron in the system can only be done via a single molecular energy state. In this regime, Fig.~\ref{Fig1} (C, D) represents in a sense a spectroscopy of single molecular levels. The irregular stability diagram therefore reflects the complexity of the molecular addition spectrum of the system as already mentioned.

In what follows we present a theoretical analysis of the experimental results using a simplified model that captures qualitatively the main features observed in the conductance maps.
\section{Model and Methods} \label{sec:theory}
The double quantum dot (DQD) system is modeled in the constant interaction model by the Hamiltonian\cite{Wiel2002} (for related models, see e.g. Ref. [\onlinecite{DasSarma2011}] and references therein)
\begin{equation} \label{eq:hamilt}
  H=H_C+ H_{t}+H_e+H_{V}+H_{el}\;.
\end{equation}
Here $H_C$ describes the electrostatic interactions
\begin{eqnarray} \label{eq:coul}
H_C &=& \sum_{\ell=d,D} \frac{U_\ell}{2} (\hat{N}_{\ell}-\mathcal{N}_{\ell} )^2 \nonumber\\&+&U_{dD}(\hat{N}_D-\mathcal{N}_D )(\hat{N}_d-\mathcal{N}_d )
\end{eqnarray}
where $\mathcal{N}_{\ell}=C_{g\ell} V_{g\ell}/U_\ell$, $C_{g\ell}$ is the capacitance of dot $\ell$ with its corresponding gate electrode, $U_\ell$ is the charging energy and $U_{dD}$ is given by the QDs mutual capacitance.\cite{Kastner92,KouwenetalRev97,AleinerBG02,Alhassid00} 
\begin{eqnarray}
H_{t}=\sum_{\sigma,\alpha,\beta}t_{dD}^{\alpha\beta}
\left(d^\dagger_{d\alpha\sigma} d_{D\beta\sigma} + h.c\right),
\end{eqnarray}
describes the tunneling coupling between the different levels on the two dots with effective single electron energy levels $\tilde{\epsilon}_{\ell\alpha}$:
\be
H_e = \sum_{\ell={d,D}}\sum_{\sigma,\alpha}\tilde{\epsilon}_{\ell\alpha} d_{\ell\alpha\sigma}^\dagger d_{\ell\alpha\sigma},
\ee
and
\be
H_V =\sum_{\alpha}\sum_{\nu=L,R} \sum_{k,\sigma}V_{k\alpha\nu}\left[c_{\nu k\sigma}^\dagger d_{d\alpha\sigma} + h.c. \right],
\ee\\
describes the coupling between QD $d$ and the left (L) and right (R) electrodes, which are modeled by two non-interacting Fermi gases:
\be
H_{\text el} = \sum_{\nu, k,\sigma} \epsilon_k c_{\nu k\sigma}^\dagger c_{\nu k\sigma}.
\ee

We follow Refs. [\onlinecite{Meir1992,Pastawski1992,Beenakker1991}] to calculate the conductance through the system,
\be\label{eq:Gsimp}
G=\frac{e^2}{\hbar}\int d\epsilon\left[-\frac{\partial f(\epsilon)}{\partial \epsilon}\right]\text{Tr}\left\{\frac{{\mathbf \Gamma^R}{\mathbf \Gamma^L}}{{\mathbf \Gamma^R}+{\mathbf \Gamma^L}}{\mathbf A}(\epsilon)\right\}.
\ee
Here $A(\omega)$ is the QD spectral density and we have assumed proportional ($\Gamma^L\propto \Gamma^R$) and  energy independent dot--leads hybridization functions:
\be \label{eq:gamma}
[{\bf \Gamma}^{L(R)})]_{\ell,\ell^\prime}= 2 \pi \rho_{L(R)}(\Ef)
V^*_{L(R),\ell}(\Ef)V_{L(R),\ell^\prime}(\Ef).
\ee
where $\Ef=0$ is the Fermi energy of the electrodes, $\rho_{L(R)}(\epsilon)$ is the electronic density
of states of the left (right) electrode, and $V_{L(R),\ell}(\eps)$ equals $V_{kL(R),\ell}$ for $\eps=\eps_k$. 

To lowest order in $\Gamma/\kt$ we replace in Eq. (\ref{eq:Gsimp}) the exact spectral density $A(\epsilon)$ of the isolated DQD:
\bea
A^\sigma_{n,m}(\epsilon)&=&\frac{1}{Z}\sum_{i,j} (e^{-\beta E_i}+e^{-\beta E_j}) \langle\Psi_j|d_{n\sigma}^\dagger|\Psi_i\rangle\nonumber\\&\times&
\langle\Psi_i|d_{m\sigma}|\Psi_j\rangle \delta[\epsilon-(E_j-E_i)],
\eea\\
where $|\Psi_i\rangle$ and $E_i$ are the exact eigenfunctions and eigenenergies of the DQD, and $Z=\sum_i e^{-\beta E_i}$ is the partition function.
We get:
\bea
G&=&\frac{e^2}{\hbar}\sum_{n,m} \Gamma_{n,m} \sum_{i,j} (P_i+P_j) \left[\left.-\frac{\partial f(\omega)}{\partial \omega}\right|_{E_i-E_j}\right]\nonumber\\&\times&\langle\Psi_j|d_{n\sigma}^\dagger|\Psi_i\rangle\langle\Psi_i|d_{m\sigma}|\Psi_j\rangle
\eea
where $P_i=e^{-\beta E_i}/Z$.

In the experimental setup, $\Gamma$ is nonzero only for the small dot $d$, and we choose it to be level independent: 
\be
\Gamma_{n,m}=\left\{\begin{matrix}\Gamma\delta_{n,m} & \text{if}\,\,n \in d\\ 0 &\text{if}\,\, n\in D \end{matrix}\right.
\ee
We finally have:\\
\bea \label{eq:condf}
G&=&\frac{e^2}{\hbar}\frac{\Gamma}{\kt} \sum_{i,j} (P_i+P_j) f(E_i-E_j)f(E_j-E_i)\nonumber\\&\times&\sum_n |\langle\Psi_j|d_{dn\sigma}^\dagger|\Psi_i\rangle|^2 
\eea
valid for $\Gamma \ll \kt$.

It is clear from this formula that the conductance is suppressed for $|E_i-E_j|\gg \kt$, i.e. the conductance is low, unless two states with $N+1$ and $N$ electrons in the DQD are nearly degenerate allowing the charge in the molecule to fluctuate.\cite{Beenakker1991} 
This is not a sufficient condition, for a charge fluctuating in and out of a state whose weight is mainly located in the large dot, the matrix elements $\sum_n|\langle\Psi_j|d_{dn\sigma}^\dagger|\Psi_i\rangle|^2$ are small and suppress the conductance.

The calculation of the conductance maps then reduces to obtaining the eigenenergies and eigenfunctions for the isolated DQD molecule. 
This can be done by exact diagonalization for a limited number of states on each dot, due to the exponential increase of the size of the Fock space with the number of levels.

\section{tunneling crossover} \label{sec:numdis}
In this section we present numerical results for the conductance maps that reproduce qualitatively the main experimental observations, as a crossover from weak to strong interdot tunneling. 
We exactly solve a model of an isolated DQD with three levels on each quantum dot and use Eq. (\ref{eq:condf}) to calculate the conductance in the weak dot-leads coupling regime ($\Gamma/\kt\ll 1$)

We use the experimentally obtained values for the parameters: U$_d$=700 $\mu$eV, $\Delta_d$=150 $\mu$eV and U$_D$=250 $\mu$eV, $\Delta_D$=20 $\mu$eV, and $U_{dD}=100\mu$eV. We consider fixed intradot level splittings and the same interdot tunneling coupling $t_{dD}^{\alpha\beta}=t_{dD}$ for all levels $\alpha,\beta$. To model the tunneling crossover observed in the experiments we include a linear crosstalk of the gate that determines the tunneling coupling between the dots with the gate voltages $V_{gd}$ and $V_{gD}$: 
\be
\td \propto V_{gt} + \alpha_D V_{gD} + \alpha_d V_{gd} 
\ee

\begin{figure}[tb]
 \centering
 \includegraphics[width=0.5\textwidth]{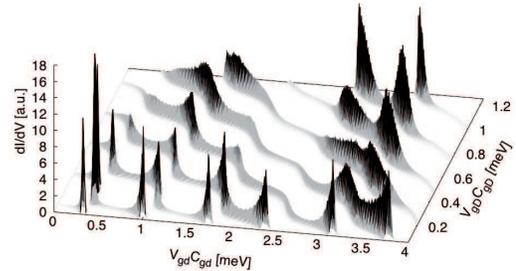}
 \caption{ Three-dimensional representation of the calculated conductance for a double quantum dot with $3\times 3$ levels, including a gate voltage dependence of $t_{dD} = 0.015$ meV$+ 0.01 $ meV$(\mathcal{N}_D+\mathcal{N}_d/3)$ and $k_BT=0.0075$ meV.
 }
\label{fig:3D}
\end{figure}

Figure \ref{fig:3D}	shows a three-dimensional representation of the calculated conductance as a function of the gate voltages including a crosstalk with the interdot tunneling amplitude. \footnote{In the experiments, a crosstalk is observed between gates $V_{gd}$ and $V_{gD}$. Assuming a linear crosstalk we have:
$\mathcal{N}_d U_d= C_{gd} V_{gd} +C_{gd}^D V_{gD}$ and
$\mathcal{N}_D U_D= C_{gD} V_{gD} +C_{gD}^d V_{gd}$.
These relations can be easily inverted to obtain the experimental axes $V_{gd}$ and $V_{gD}$. 
The result is a linear transformation that may include scaling, shear and rotation. 
This type of transformations need to be considered to obtain a more quantitative agreement between theory and experiment.
} 
 The hopping amplitude increases linearly from a minimum at the lower left corner of the figure to its ma\-xi\-mum at the top right corner. As we will see in the next section, the system 
goes from a low-$\td$ regime for small $\mathcal{N}_D, \mathcal{N}_d$, to a high-$\td$ regime for high $\mathcal{N}_D, \mathcal{N}_d$, both characterized by a regular array of conductance peaks (ho\-ney\-comb diagram), associated to the charging of the small QD.  In the intermediate tunneling regime, the electronic wavefunctions are highly delocalized between the two QDs and the DQD enters a single-dot molecular regime where the conductance maps present diagonal lines of high and relatively uniform conductance. 
The result is an apparent merging and crossing of peaks with increasing $\ND$ as observed experimentally.
Eventually, the emergence of a regular pattern of peaks at high $\td$ implies that the wavefunctions are localized on each dot as in the low-$\td$ case. As we will show in the next section this is a consequence of the structure of the wavefunctions in the regime $\Delta_d\sim\td\gg\Delta_D$.
 
\begin{figure}[tb]
 \centering
 \includegraphics[width=0.5\textwidth]{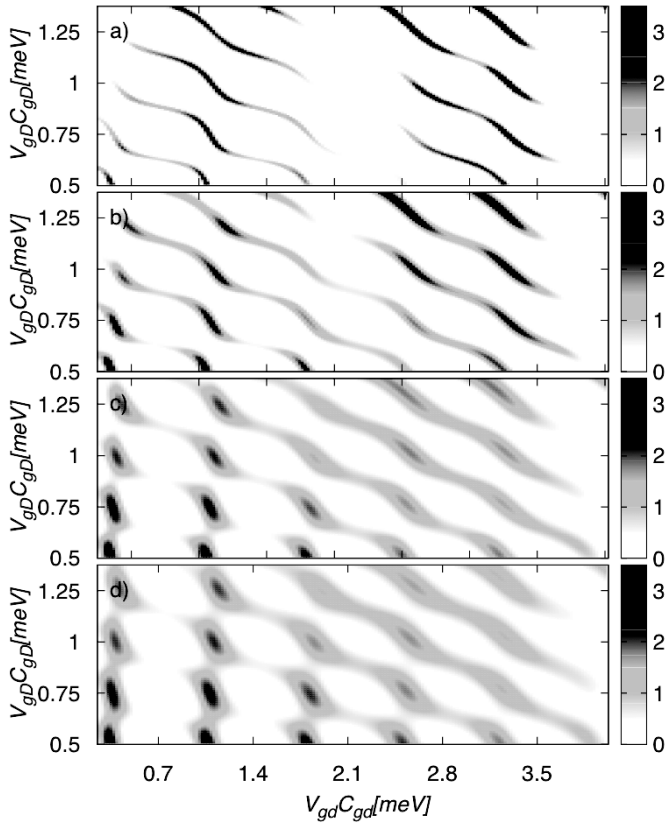}
 \caption{ Calculated conductance maps for a double quantum dot with $3\times 3$ levels, including a gate voltage dependence of $t_{dD} = 0.015 \meV + 0.01 \meV (\mathcal{N}_D+\mathcal{N}_d/3)$ and different values of the temperature: a) $k_BT=0.0075\meV$, b) $k_BT=0.02\meV$, c) $k_BT=0.04\meV$, and d) $k_BT=0.05\meV$. Other parameters are $U_d=0.7\meV$, $U_D=0.25\meV$, $U_{dD}=0.1\meV$, $\Delta_d=0.15\meV$, and $\Delta_D=0.02\meV$.
 }
\label{fig:tT}
\end{figure}

We now focus on the effect of the temperature on the conductance maps. 
Figure \ref{fig:tT} presents conductance maps calculated with the same parameters as in Fig. \ref{fig:3D} for different values of the temperature. Each panel is calculated using a different temperature but the parameters are otherwise equal. 
Increasing the temperature allows us to observe its effect on regions with different values of the hopping amplitude and investigate the {\it regularization} of the patterns observed experimentally. 

For temperature regimes where $\kt$ is much smaller than the DQD's energy level spacings, the main effect of increasing the temperature is to increase the width and reduce the height of the Coulomb blockade peaks. This can be readily seen from Eq. (\ref{eq:condf}) assuming that a single state from each charge sector contributes to the conductance at a given peak. 
When the temperature becomes of the order or larger than the level spacing in a given charge sector, several states may contribute to the conductance producing in some cases a qualitative change in the conductance maps. Such changes are expected in the present DQD geometry whenever the states that contribute to a single CB peak have a markedly different weight on each dot. 
In that case, the intensity of the conductance peak at low temperatures will be very different depending on the nature of the state that dominates the charging: a small conductance if the large dot is being charged and a large conductance if the small dot is being charged. At temperatures larger than the level spacing, however, several states with a different weight on each dot may statistically contribute to the charging resulting in an intermediate value of the conductance.

This type of behavior is observed in Fig. \ref{fig:tT}, where an increase in the temperature produces a broadening of the charge degeneracy lines and leads to a more homogeneous intensity of the conductance along them. 

At the highest temperature shown in the lowest panel of Fig. \ref{fig:tT}, the conductance pattern presents a regular lattice of maxima. 
This is the expected behavior for $\kt\gtrsim \td$ with the position of the conductance maxima given by the charging energies as in the $\td \to 0$ case. 

\section{Tunneling Regimes}\label{sec:regimes}
In the previous section we showed that the main features of the measured conductance maps can be reproduced numerically and that the observed merging and apparent crossing of peaks are associated to a crossover from weak to strong interdot tunneling regimes. In this section we characterize the different tunneling regimes. 

We start with the case of uniform charging energies ($U_{dD}=U_d=U_D=U$) that is obtained in the regime of large interdot capacitance. This case allows for an analytical solution and already contains the underlying structure of the general case.

\subsection{Large interdot capacitance}\label{ssec:UUU}
For $U_{dD}=U_d=U_D=U$ the isolated DQD can be readily solved as the interaction term only depends on the total number of electrons in the molecule $N=N_d+N_D$ which is a good quantum number. The energy of the system is given by $E^N =  E_C^N + \sum_\alpha^N \varepsilon_\alpha$, where the $\varepsilon_\alpha$ are the eigenenergies of $H_{tb}=H_t+H_e$, and $E_C^N=\frac{U}{2}(N-\mathcal{N})^2$, where $\mathcal{N}=\mathcal{N}_d + \mathcal{N}_D$. The charge degeneracy points ($E_j-E_i=0$) satisfy the equation
\be
\mathcal{N}_d+\mathcal{N}_D = N+\frac{1}{2} + \varepsilon_{N+1}/U
\ee
that determines a series of parallel lines in the $(\mathcal{N}_d,\mathcal{N}_D)$ plane. The intensity associated to these lines in the conductance map is proportional to the weight of the additional electron's wavefunction in dot $d$. 
In the case of uncoupled dots ($t_{dD}=0$), only the lines associated to the charging of dot $d$ present a maximum in the conductance.
In the general case ($t_{dD}\neq 0$) the wavefunctions of the DQD are delocalized between the dots and the intensity of the conductance lines is modulated accordingly. Surprisingly,  for large $\td$ the wavefunctions are again mainly localized on each dot as in the small $\td$ regime.

To show this latter point, we further simplify the model by considering $\tilde{\varepsilon}_{\ell\alpha}=\tilde{\varepsilon}_{\ell}$  (i.e. $\Delta_d=\Delta_D=0$) and $t^{\alpha\beta}_{dD}=t_{dD}$. Then 
\be
H_{tb}=
\left(
\begin{array}{cc}
\tilde{\varepsilon}_d\mathbf{1}&{\bf T}\\
{\bf T}&\tilde{\varepsilon}_D\mathbf{1}
\end{array}
\right),
\ee
where $\mathbf{1}$ is the identity matrix and
\be
{\bf T}=
\left(
\begin{array}{cccc}
t_{dD}&t_{dD}&\cdots&t_{dD}\\
t_{dD}&\cdots&\cdots&t_{dD}\\
\cdots&\cdots&\cdots&\cdots\\
t_{dD}&\cdots&\cdots&t_{dD}
\end{array}
\right).
\ee
The Hamiltonian matrix $H_{tb}$ can be diagonalized exactly for its eigenvectors. Among all wavefunctions, only two are strongly delocalized between the dots, having half of the weight on each dot, the rest of the states are either fully localized at dot $d$ and have energy $\tilde{\varepsilon}_d$ or are fully localized at dot $D$ and have energy $\tilde{\varepsilon}_D$ (assuming $\tilde{\varepsilon}_d\neq\tilde{\varepsilon}_D$). For a finite level spacing on each dot $\Delta_D,\Delta_d\ll t_{dD}$, the states remain localized in one of the dots, with only a small weight ($\ll 1$) on the other dot. This structure of eigenstates persists even if one of the level spacings becomes of the order or even larger than the hopping amplitude, e.g. $\Delta_d \gtrsim t_{dD}\gg \Delta_D$.

\subsection{Experimental situation: $U_d>U_D>U_{dD}$}
In the experimental situation, there is a hierarchy of interactions: $U_d>U_D>U_{dD}$ and it is generally no longer possible to solve the interaction and tight-binding parts of the Hamiltonian independently as it was done in the previous section. However, as we shall see, the analysis presented above serves as a guide when tackling the general case.  

\begin{figure}[tb]
 \centering
 \includegraphics[width=0.5\textwidth]{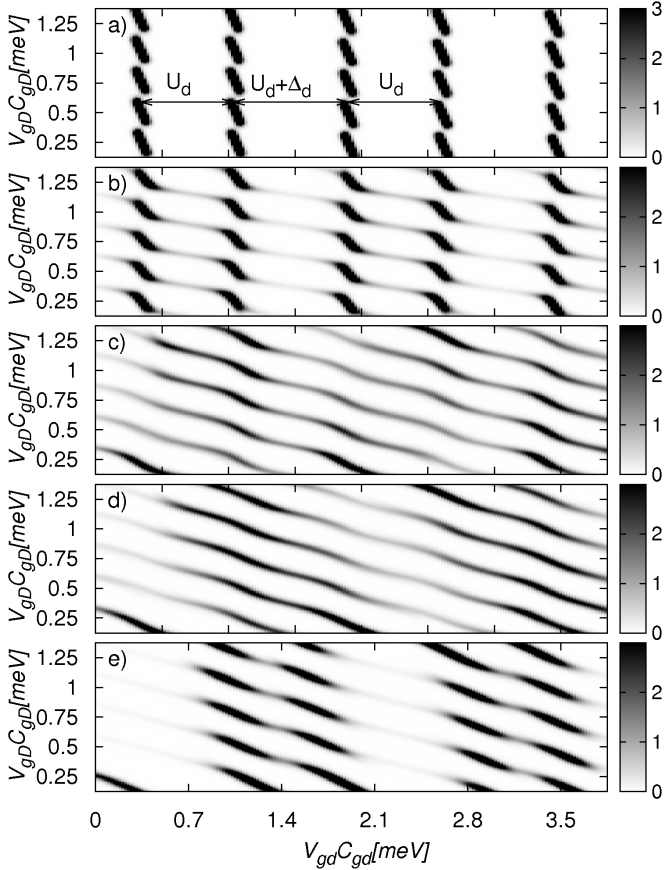}
 \caption{Calculated conductance maps for a double quantum dot with $3\times 3$ levels, for different values of the interdot hopping $\td$ a)$0\meV$, b)$0.02\meV$, c)$0.04\meV$, d)$0.05\meV$, and e)$0.08\meV$. Other parameters are $U_d=0.7\meV$, $U_D=0.25\meV$, $U_{dD}=0.1\meV$, $\Delta_d=0.15\meV$, and $\Delta_D=0.02\meV$. The high-$t_{dD}$ and low-$\td$ with segments of high conductance are clearly observed. }
\label{fig:tchange}
\end{figure}

We first focus on the limit of small $t_{dD}$ where the charge in each dot is well defined. The charging conditions for dots $d$ and $D$ are given by
\bea \label{eq:CDL}
\nonumber
\mathcal{N}_d = N_d+\frac{1}{2} + \frac{\tilde{\varepsilon}_{dN_d+1}}{U_d} +  \frac{U_{dD}}{U_d}(N_D-\mathcal{N}_{D}),\\
\mathcal{N}_D = N_D+\frac{1}{2} + \frac{\tilde{\varepsilon}_{DN_D+1}}{U_D} +  \frac{ U_{dD}}{U_D}(N_d-\mathcal{N}_{d}),
\eea
respectively, where $\tilde{\varepsilon}_{\ell N_\ell+1}$ is the energy of the concerned level on dot $\ell$,
and determine two sets of parallel straight lines in the $(\mathcal{N}_d,\mathcal{N}_D)$ plane. The charging of the DQD and the conductance maps are determined by these equations and the result, for $U_{dD}>0$, is a series of high conductance segments with slope $-U_{d}/U_{dD}$ associated to the charging of the small dot [see Fig. \ref{fig:tchange}(a)]. The endpoints of these segments are given by the intersections of the charge degeneracy lines (CDL) of the small dot by the CDLs of the large dot, i.e by triple degeneracy points: $E(N_d,N_D)=E(N_d,N_D+1)=E(N_d+1,N_D)$ and $E(N_d,N_D)=E(N_d+1,N_D-1)=E(N_d+1,N_D)$. The separation between two consecutive segments, for a fixed $\mathcal{N}_D$, is associated to the extra energy required to add an electron and depends on the parity of electron number on dot $d$: it is $U_d$, when a second electron is added to a partially occupied level (odd electron valley) and $U_d+\Delta_d$ when it is added to an empty level (even electron valley). 

A finite but small $\td < \Delta_D, \Delta_d$ produces a distortion of the high conductance segments, and a small peak in the conductance associated to the charging of the large dot, due to interdot mixing. The main features in this regime can be understood within a simplified model with a single level on each dot [see Fig. \ref{fig:tchange}(b)].

For intermediate values of $\td$ the states are strongly mixed between the two dots and the charge on each dot is no longer well defined. In this case, the conductance maps are not expected to have a regular pattern of segments with high conductance. Instead, as is shown in Figs. \ref{fig:tchange}(c) and \ref{fig:tchange}(d), the situation is similar to the high interdot capacitance case (see previous section) with the conductance map showing diagonal lines of high conductance separated by an effective interaction. 

In the regime of large $\td$, however, the state wavefunctions are, as in the uncoupled case, mainly localized on each dot.  Equations (\ref{eq:CDL}) determine the conductance map patterns in the limit $t_{dD}\to 0$ but also for $t_{dD}\to \infty$ and give their qualitative shapes in the regimes  $t_{dD}\ll \Delta_D$ and $\Delta_d \sim t_{dD}\gg\Delta_D$. 

An important difference between the large and small $\td$ limits is that while in the small $\td$ limit, {\it all} states have a well defined number of electrons on each dot (excluding specific regions in the parameter space), in the large $\td$ limit there are only two states which are strongly delocalized between the dots. 

One of these states, which is fully symmetric between dots, is the ground state and is the first to be charged as the gate voltages are swept. The complete charging of this state adds a single electron to each dot and therefore {\it changes the parity} for the subsequent charging of the dots. This parity change alters the sequence of distances between high conductance peaks as the gate voltage of the small dot is swept: the separation between two consecutive segments, for a fixed $\mathcal{N}_D$, is now: $U_d$ when adding an {\it odd} electron on dot $d$ and $U_d+\Delta_d$ for an {\it even} electron. 

The above-mentioned parity effect persists even for $\td\lesssim \Delta_d$, and the pattern of peaks in the conductance follows qualitatively what is expected in the high $\td$ limit. 
A qualitative understanding of the peaks positions and shapes can be obtained considering that for a finite but large value of $\td$ the states are not {\it fully} localized on each dot. A small interdot mixing of the states reduces the effective charging energy of dot $d$ and increases the effective interdot interaction. This leads to a reduction of the slope of the high conductance segments [see Fig. \ref{fig:tchange}e)] which is given by $-U_{d}/U_{dD}$ in the $\td\to 0$ and $\td\to \infty$ limits.

\section{Summary and conclusions} \label{sec:concl}
We have studied the transport through a double-quantum-dot system in a side-coupled configuration as a function of temperature and interdot tunneling coupling. We have focused on the weak QD-electrodes coupling regime and analyzed the structure of the DQD´s molecular wavefunctions. The topology of the device allows us to study via transport measurements, how the wavefunction weight is distributed between the QDs for each molecular state. The geometry of the device makes it possible to explore different tunneling coupling and temperature regimes. We have constructed and solved a simplified model that reproduces the experimentally observed regimes.  

For a weak interdot coupling ($\td\ll\Delta_D,\Delta_d$) the molecular states can be accurately described considering a model with two-levels, one from each QD, coupled by a hopping term $\td$.
The resulting molecular wavefunctions are essentially localized on one of the quantum dots for most values of the plunger gate voltages. The conductance in the ($V_{gd},V_{gD}$) plane reflects this structure of molecular eigenstates and shows a series of CB peaks associated to the charging of the QD directly coupled to the electrodes.
Much weaker CB peaks are obtained as the side-coupled QD is charged due to a small mixing between the QDs.

When the interdot coupling is increased, more le\-vels from each QD are involved in the formation of a given molecular state. An intermediate tunneling regime ($\td\gtrsim\Delta_D$) can be reached where the molecular wavefunctions are strongly delocalized between the QDs. In this situation, the conductance maps present a series of lines of high and relatively uniform conductance in the ($V_{gd},V_{gD}$) plane and resemble those expected for a {\it single} quantum dot coupled to two plunger gates.
 
For large enough interdot coupling,($\td\sim \Delta_d\gg\Delta_D$), the nature of the eigenfunctions changes and most molecular states become increasingly localized on each QD as in the weak $\td$ limit. There is, however, an important difference between these two regimes due to the emergence, in the high $\td$ limit, of two states of a different nature that result from a mixing of several levels from each dot.
These states are a symmetric and antisymmetric combination between the states of the two QDs, and have (for $\td>0$), a lower and higher energy than their component states, respectively. The charging of the lowest lying of these states involves adding a single electron to each QD and alters the even-odd sequence of CB peaks producing a shift of the high conductance peaks in the ($V_{gd},V_{gD}$) plane.

In the experiments there is a crosstalk between the plunger gates of each QD and the gate controlling the tunneling barrier between the QD. As a consequence, different regions of the ($V_{gd},V_{gD}$) plane have associated different intensities of tunneling coupling and it is possible to observe the above-mentioned regimes in a single conductance map. The crossover between the different regimes gives rise to a complex evolution of the CB peaks with mergings and apparent crossings.

Finally, we analyzed the effect of the temperature on the transport properties. For $\kt>\td$, several levels contribute to the conductance of each Coulomb blockade peak, the pattern of conductance maxima is similar to that of weakly coupled dots in the $\kt\sim \td$ regime, and resemble those of uncoupled dots ($\td\to 0$) for $\kt\gg \td$.

\begin{acknowledgments}
We thank P. Simon for driving our interest  towards such a double dot setup, and S. Florens for stimula\-ting discussions. P.S.C., G.U. and C.A.B. acknowledge financial support from PIP 11220080101821 of CONICET and PICT-Bicentenario 2010-1060 of the ANPCyT. T. M. acknowledges partial funding from Marie Curie ERG 224786. D.M., C.B. and L.S. acknowledge financial support from ANR project ``QuSpin''. D.F. acknowledges support from PICS 5755 of CNRS.
\end{acknowledgments}

\bibliographystyle{apsrev4-1}

\bibliography{references}

\end{document}